\documentclass[12pt]{article}
\usepackage{graphicx}
\usepackage{epsf}

\makeatletter
\@addtoreset{equation}{section}
\makeatother

 \def\bd{\begin{document}} \def\ed{\end{document}}
\def\ds{\documentstyle} \let\fr=\frac \let\bl=\bigl \let\br=\bigr
\let\Br=\Bigr \let\Bl=\Bigl
\let\bm=\bibitem
\let\na=\nabla
\let\pa=\partial \let\ov=\overline
\newcommand{\be}{\begin{equation}}
\newcommand{\ee}{\end{equation}}
\def\ba{\begin{array}}
\def\ea{\end{array}}
\newcommand{\ho}[1]{$\, ^{#1}$}
\newcommand{\hoch}[1]{$\, ^{#1}$}
\newcommand{\bea}{\begin{eqnarray}}
\newcommand{\eea}{\end{eqnarray}}
\newcommand{\ra}{\rightarrow}
\newcommand{\lra}{\longrightarrow}
\newcommand{\Lra}{\Leftrightarrow}
\newcommand{\ap}{\alpha^\prime}
\newcommand{\bp}{\tilde \beta^\prime}
\newcommand{\tr}{{\rm tr} }
\newcommand{\Tr}{{\rm Tr} }
\newcommand{\NP}{Nucl. Phys. }
\ifx\ltimes\undefined
\newcommand{\ltimes}{{\kern3pt\hbox{\vrule width 0.4pt height 5.30pt depth
.0pt}\kern-1.76pt\times\kern1pt}}
\fi

\newcommand{\tamphys}{\it The Blackett Laboratory, Imperial College London,\\ 
Prince Consort Road, London SW7 2AZ\\}

\newcommand{\auth}{M. J. Duff }

\begin{document}
\begin{flushright}
\hfill{Imperial/TP/2021/mjd/2}\\

\end{flushright}
\vspace{10pt}

\begin{center}
{ \large {\bf Chris Isham: mentor, colleague, friend.}}

\vspace{20pt}

\auth

\vspace{10pt}

{\tamphys}

\vspace{20pt}

\underline{ABSTRACT}

\end{center}

Celebrating fifty years of collaboration and friendship with Chris Isham.

\vfill
\leftline{}

\vfill
\leftline{}

\newpage




\section{P. T. Matthews}
It was the day man landed on the moon. 
I received a telegram at my parents' home 
in Manchester asking me to phone 
Prof. P. T Matthews at Imperial College 
London. In common with many British 
working class households in 1969, my  
parents did not own a telephone and I had 
to use a public phone box. Imperial College had a 
formidable reputation for excellence. 
They had rejected my application for 
an undergraduate degree in Physics three
years earlier and I was lucky to attend Queen Mary College 
instead. Having squandered my time in the 
sixth form at school,  Imperial's decision was 
totally justified but I worked  hard at QMC and earned a place in the Part III 
Mathematical Tripos at Cambridge, a stepping 
stone to a PhD. So as I put the coins in the phone-box, I wondered why the call from Imperial?
My supervisor at QMC, John Charap, was previously at Imperial. Perhaps he had put in a good word?
 
Indeed Matthews dismissed Part III as inconsequential
and said I would be much better off  
at Imperial where I could start my PhD straight
away. True, there were lectures in the first year 
leading to a DIC (Diploma of Imperial College,
roughly equivalent to a Masters Degree) but 
my place in the PhD programme was 
guaranteed unlike at Cambridge where 
I faced another year of swotting for Part III exams 
before starting a doctorate. That settled it. 
 Not only would I achieve 
my ambition of becoming Dr. Duff by the age of 23 (having left school a year early), but no more exams!

However, when I arrived for my first day at 
Imperial, Matthews, who was then Head of the 
Physics Department said ``You  
contact me only if you need bailing out of jail''.
He was only half joking as he had that  
very morning bailed out the President 
of the Imperial College Student Union, 
who had been arrested at an anti-Vietnam-war 
demonstration. He was Piers Corbyn, whose 
elder brother Jeremy Corbyn later became 
Leader of the Labour Party.  Matthews was himself something of a celebrity
 in the anti-war movement, frequently photographed 
with the actress Vanessa Redgrave\footnote{Just this week Vanessa Redgrave and Piers Corbyn were again in the news:  Redgrave was made a Dame in the New Year's honours list and  Corbyn was arrested.}.  Anyway he followed up his comment 
with ``So go down the corridor and knock on Salam's door; he 
will allocate your PhD project.''
\begin{figure}\label{fig:p1}  
\begin{center}
\includegraphics[scale=0.5]{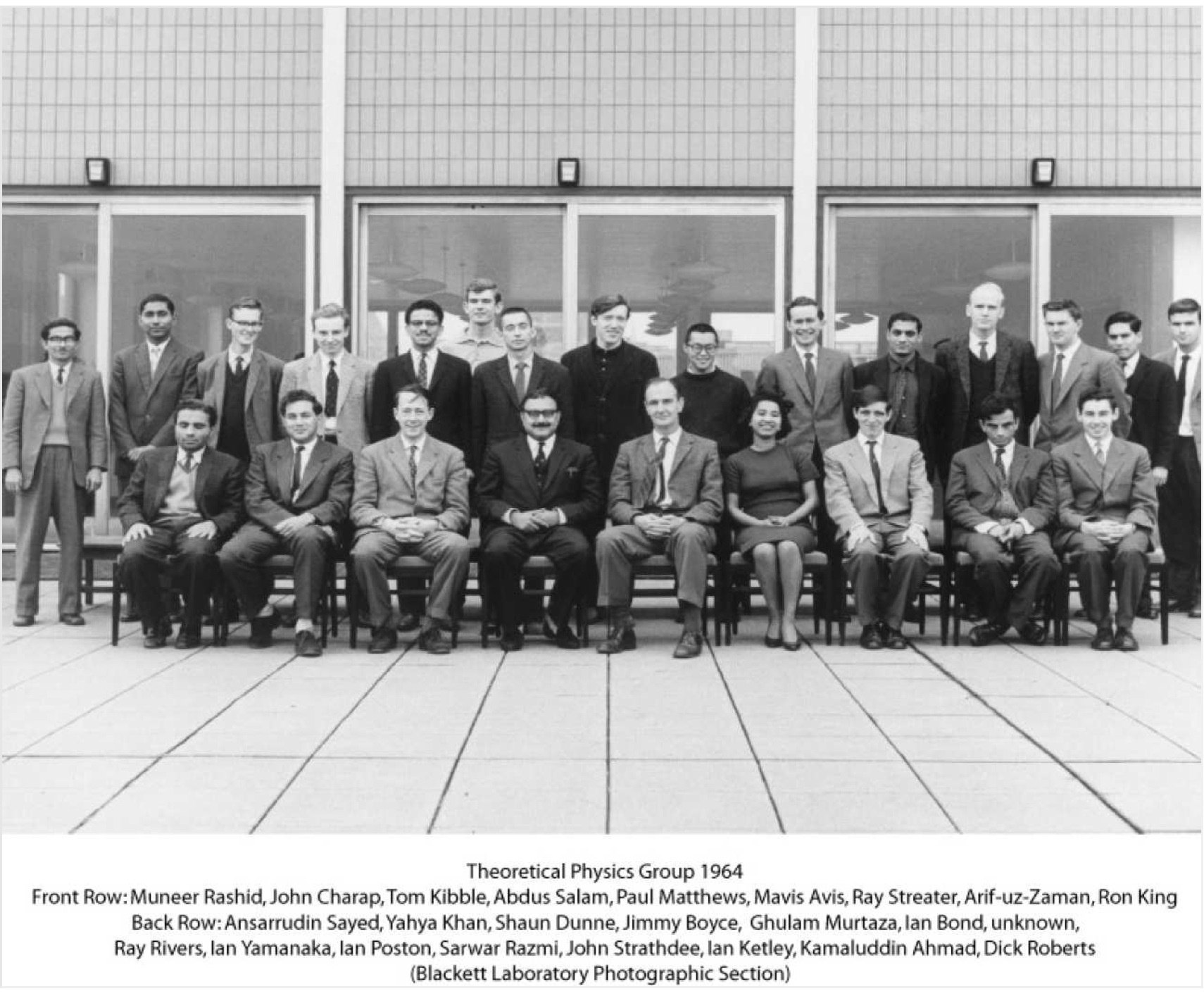}
\caption{\footnotesize{Four of the characters in our story: Paul Matthews, Abdus Salam, John Charap, and John Strathdee, may be found in this 1964 photograph of the Imperial College Theoretical Physics Group.}}
\end{center}
\end{figure}

\section{Abdus Salam}
Now even  as a schoolboy in Manchester, the name of 
Abdus Salam was familiar to me and to my
scientifically-minded friends as a leading light in 
theoretical physics. By 1969 he had already 
written the papers that would earn him the 
Nobel Prize (shared with Glashow and Weinberg) ten years later. Although it was a warm day, Salam 
had on a three-bar electric heater and wore one of those college 
scarves that were fashionable at the time.
``My work up until now has been focussed on three of the four 
fundamental forces: strong, weak and electromagnetic.'' Salam said.
``It is time to include gravity: an astonishing force.''

As I was to discover, Salam rarely had just one idea for students to work on but several.
My memory may be unreliable but I think at that first meeting there were three: (1) Infinity suppression using non-polynomial Lagrangians, especially gravity (2) {\it f-g theory}, a theory with both a massless spin 2 graviton and a massive spin 2 f-meson (3) Generating the Schwarzschild black hole metric using Feynman diagrams. Moreover, it was impossible to tell which of his ideas should be taken seriously and which should be consigned to the dustbin of history. I think my first publication \cite{Duff:1971rjh}, based on (1) ``The $\pi^+-\pi^0$ mass-difference using gravity-modified hadron electrodynamics'', belongs in the second category. For more on Salam, see \cite{Duff:2008zz}.

\section{Chris Isham}

In that era Salam was often away on his travels, especially to the International Centre for Theoretical Physics in Trieste (ICTP) which he founded in 1964. He frequently worked with three collaborators: John Strathdee based in Trieste, Bob Delbourgo based at Imperial and the wunderkind  Chris Isham who was appointed to a permanent  position at Imperial immediately after his PhD, which was virtually unheard-of. However, he spent the first year in Trieste while mine was in London, so I knew him only by reputation. And what a reputation it was! His mathematical prowess was legendary. The kind of maths he was doing, involving group theory, geometry  and topology is now commonplace in theoretical high energy physics but was way ahead of its time in 1969. His admirers included other forward-thinking physicists such as Dennis Sciama and Roger Penrose in Oxford and  Bryce Dewitt, Stanley Deser and Roman Jackiw in the USA.

    As Salam continued to travel, it was to Chris that I turned to on his return to Imperial for help with my PhD. On the purely academic side I fully expected  and received superb supervision. What I had not anticipated  was how much fun it turned out to be. Chris had an infectious sense of humour and we soon became great friends. He and his wife Valerie frequently invited me for dinner at their home in Rayners Lane. A perusal of the London Underground Map will confirm that Rayners Lane is on the boundary rather than in the bulk, and Chris did a lot of his work on the train travelling from and to {\it Metroland} as Poet Laureate John Betjeman called that sprawling suburbia\footnote{See https://en.wikipedia.org/wiki/Metro-land for earlier history.}. We also enjoyed a wonderful summer at the ICTP staying in John Strathdee's flat in the heart of Trieste, that most haunting and melancholy town to which I would return in 1972 for my first postdoc.  The aroma of spaghetti alle vongole in Viale XX Settembre lingers on.

    It was not all good news, however. Chris was diagnosed with a degenerative disease which resembled Muscular Dystrophy but whose exact diagnosis was to baffle the top minds in medicine in the years to come. Chris was stoic but this debilitating illness  was to rob him of the glittering prizes \footnote{A notable exception being the 2011 Dirac Gold Medal and Prize of the Institute of Physics.} that would surely have come his way had he remained in good health. Perhaps, as a Gnostic, he drew comfort from his faith.

Chris and I spent our first months trying unsuccessfully to find exact solutions to f-g theory \cite{Isham:1971gm},  the second  item in Salam's list. There were in any case arguments that  massive spin 2 should also be consigned to the dustbin. When interacting with gravity there is an unphysical degree of freedom; the Boulware-Deser \cite{Boulware:1972yco} ghost and related problems with propagation \cite{Deser:2015wta}. But, and here is another lesson regarding Salam's ideas, fast forward forty  years and theory of massive gravity \cite{deRham:2010kj}  appears which avoids the ghost and opens up new avenues of research. Two of the three authors, Claudia de Rham and Andrew Tolley are, appropriately, now members of the Imperial Theory Group.

\begin{figure}\label{fig:p2}  
\begin{center}
\includegraphics[scale=0.6]{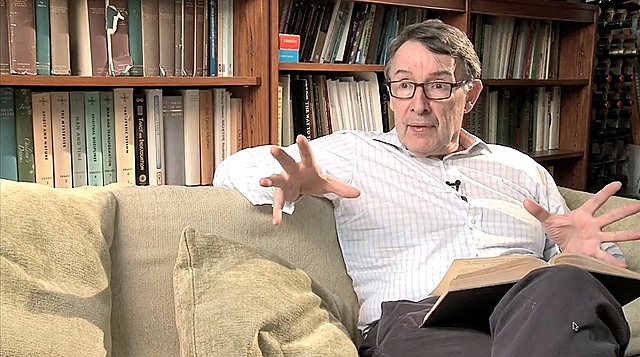}
\caption{\footnotesize{Chris in typical animated mood} }
\end{center}
\end{figure}
I tackled the third of Salam's ideas on my own as a topic for my PhD thesis but Chris was always on hand to lend his advice. This was a fascinating problem but I think the results I ended up with differed in several respects from what what Salam had in mind \cite{Duff:1973,Duff:1973zz}. To quote Salam \cite{Salam:1974zw}:
{\it Just to be provocative, let me remind Professors Wheeler \footnote{ Professor John Wheeler, former supervisor of Richard Feynman at Princeton, and the man who coined the phrase ``black hole'',} and Penrose \footnote{Sir Roger Penrose, Rouse Ball Professor of Mathematics at Oxford, who proved that Einstein's theory predicts a spacetime singularity at the center of a black hole. He was awarded the 2021 Nobel Prize for Physics for ``showing the formation of black holes must be seen as a natural process in the development of the universe.''} of a wager we had in 1971 at a lunch in a Strand restaurant hosted in absentia by Professor Bondi \footnote{Sir Herman Bondi, Professor of Mathematics at King's College London, pioneer of gravitational waves.} The bet from their side was that Duff could not possibly recover the Schwarzschild solution and its singularities by summing a perturbation series of Feynman diagrams. They were bound to lose their bet.},
but I am not sure if Wheeler, Penrose and Bondi would have agreed with Salam on the exact terms of the wager.  In any event, I saw the main purpose of the tree-graph paper as paving the way for the follow-up paper on $1/r^3$ loop corrections to Schwarzschild \cite{Duff:1973,Duff:1974ud}.

 \begin{figure}\label{fig:p4}  
\begin{center}
\includegraphics[scale=0.15]{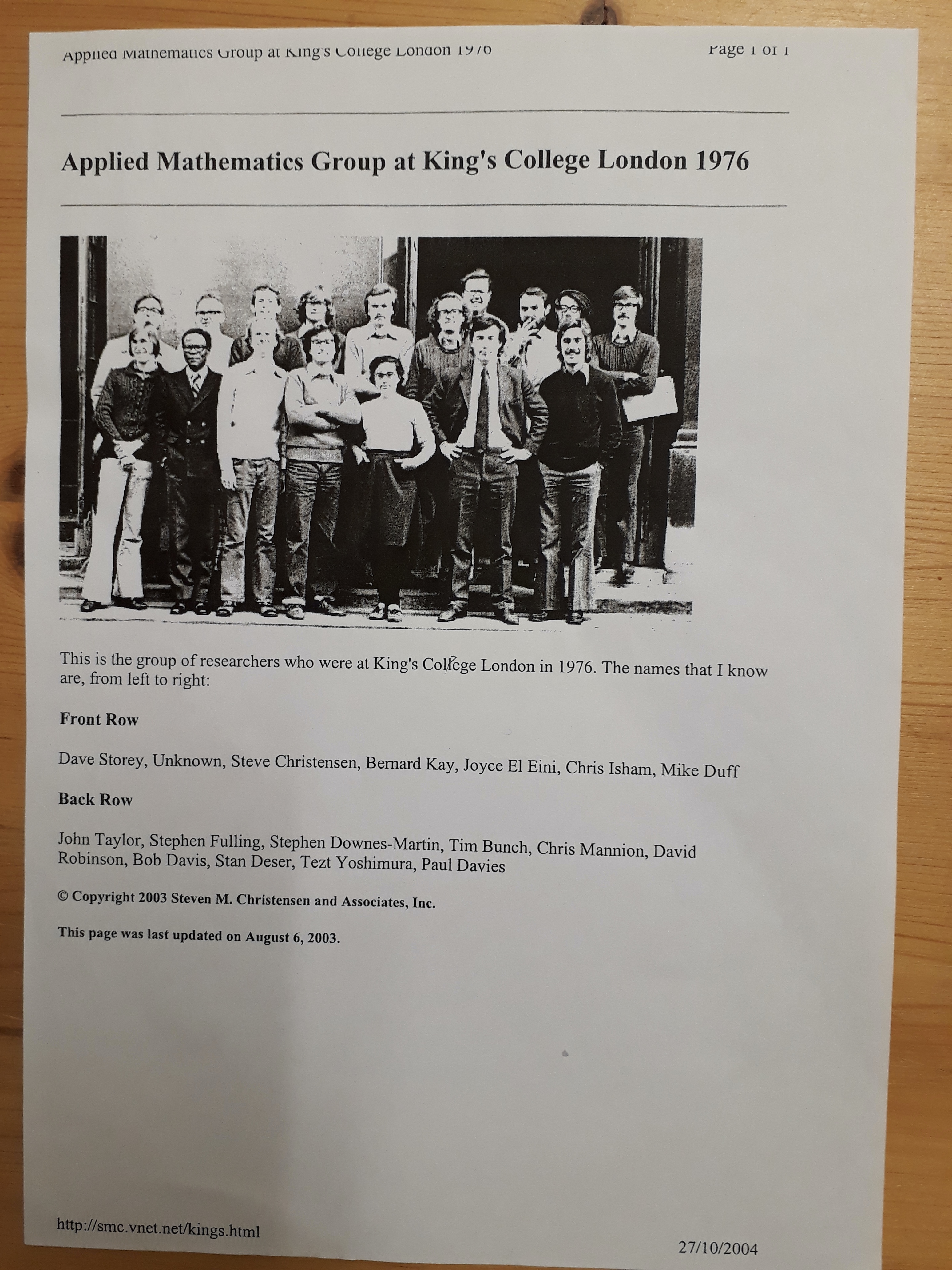}
\caption{\footnotesize{King's College London 1976} }
\end{center}
\end{figure}

In 1973 Chris took up a Readership in the Maths Department at King's College London and, after postdocs in Trieste and Oxford, I accepted his invitation to join him there on a  Science Research Council Fellowship. It was at  King's that our collaboration began in earnest, with papers on soliton solutions of non-linear sigma models and their form-factor interpretation \cite{Duff:1976av,Duff:1976td}.  The combination of permanent staff, postdocs and visitors made King's a centre of activity for classical gravity (Felix Pirani and David Robinson), quantum field theory in curved spacetime (Tim Bunch, Steve Christensen, Paul  Davies, Larry Ford, Steve Fulling and Bill Unruh) and quantum gravity (Stanley Deser, Chris Isham, John G. Taylor and myself).  Many appear in the photograph Fig. \ref{fig:p4}, courtesy of Steve Christensen.  See \cite{Robinson:2018oax} for a history.  But the visitor who had the greatest impact on Chris and me was Stanley Deser (the one smoking the cigar): non-local conformal anomalies \cite{Deser:1976yx}, solitons of chiral theories in three space dimensions \cite{Deser:1976qe}, sine-Gordon/Thirring equivalence in curved spacetime \cite{Deser:1977ue}.
I have related the Weyl anomaly saga elsewhere \cite{Duff:2020dqb}. The chiral solitons paper introduced fourth order terms in the lagrangian necessary to circumvent the Derrick no-go theorem in three dimensions\footnote{An alternative circumvention, teetering on the rim of the dustbin, was to raise the sigma-model lagrangian to the power 3/2.}.  We obtained what we considered to be nice results only to discover that some had been anticipated by Skyrme in the early 60s, although this is overlooked  in the received history of  the Skrymion revival \cite{Aitchison:2019hzj}. As for the well-known equivalence of the sine-Gordon and Thirring models, we originally claimed that it failed in curved spacetime and submitted a paper for publication only to realize they were in fact equivalent. (If my memory serves me correctly, this involved some subtleties with the beta functions and the two-dimensional Weyl anomaly).  We issued a hasty retraction but, for reasons I cannot recall, nothing in the end got published.

Somehow, in that era before universities became obsessed with profit, before constant time-consuming evaluations, before ``metrics'', ``stake-holders'' and ``end-users'', doing research was great fun and I look back on those days  of DDI collaboration with a fond nostalgia.  I must, however, recall  one unfortunate episode.  As Head of Group, John G Taylor was Principal Investigator on the Science Research Council (SRC) Grant which, amongst other things, was to provide my postdoc salary after 1976 when my Fellowship came to an end. We were verbally assured by SRC that its approval was just a formality, so I did not bother to apply elsewhere.  What happened next  must be unique in the annals of research council funding. John Taylor (who, by the way, was always very kind and pleasant to me) succumbed to the belief that the showbiz trickster Uri Geller could really bend spoons as if by magic\footnote{A view shared by the CIA. \\https://news.sky.com/story/psychic-spoon-bender-uri-geller-convinced-cia-10734232}. He appeared on a TV show supporting Geller's claims, but who should the show invite as the voice of reason to counter them but Sir Sam Edwards, chairman of SRC. Alas it was not the scientifically objective debate one might have wished for, but instead degenerated into a slanging match. A few days later the SRC grant was cancelled and I was out of a job. 

The reason for the cancellation was never spelled out by SRC but the people at King's drew their own conclusions. For a few weeks it looked as if spoon-bending had brought my career in physics  to an end until my friend John Charap found money for a one-year postdoc at QMC for which I am eternally grateful. That year was followed by two at Brandeis courtesy of Stanley Deser. As I was moving from London to Boston, Kelly Stelle was moving in the opposite direction, to join Chris at King's, and we swapped apartments.

 In 1979 I returned to Imperial College on an SRC Advanced Fellowship which evolved into a permanent Lectureship. By this time Chris had also returned as a full Professor and we renewed our collaboration. First we computed the two terms in the Euler-Heisenberg effective lagrangian  ${F_+}^2{F_-}^2$ and ${F_+}^4+{F_-}^4$, which describes the scattering of light by light, and showed that in the supersymmetric case only the first appears which may also be understood from helicity conservation. In Minkowski signature the complex fields $F_\pm=\frac{1}{2}(F \pm F* )$ correspond to off-diagonal matrix elements: vacuum to one-particle states of definite helicity \cite{Duff:1979bk}. In the non-abelian case the one-particle states are promoted to coherent states \cite{Duff:1979dy}.  The DDI collaboration also continued at Imperial with a paper on gravitational CP effects via a $\theta \epsilon^{\mu\nu\rho\sigma} R_{\mu\nu}{}^{a b}R_{\rho\sigma a b}$ term in the action \cite{Deser:1980kc}.

In 1981 it was the turn of Imperial College to host the annual Nuffield Quantum Gravity Workshop, which Chris and I organized. We made the mistake of using the in-house caterers for the Banquet:  a total disaster! Even today I meet participants who have long forgotten the physics but who remind me of the bouncing desserts. In fact the conference was not without its highlights \cite{Duff:2012sqa}.

After moving to the USA  in 1988 I saw less of Chris and even though I returned to Imperial in 2005,  our interests had diverged as I followed Supergravity, String and M-theory and he was to delve deeper into the philosophy of quantum mechanics. Fortunately, however, Chris introduced me to Leron Borsten, a very bright student of his who was doing an undergraduate project on quantum information theory.  He was to become my PhD student, collaborator and friend. In recent years Chris's health continued its inexorable decline and he came into college less and less. But I am honoured that office number 510 in the Huxley building at Imperial still bears the names of two emeritus professors: Christopher Isham and Michael Duff. 
\section{Acknowledgements}
Thanks to Stanley Deser and Leron Borsten.

\bibliography{Isham}
\end{document}